\documentclass[pre,twocolumn,amsmath,amssymb,floatfix,showpacs]{revtex4}

\usepackage{graphicx}

\begin{document}

\title{Time-Varying Priority Queuing Models for Human Dynamics}
\author{Hang-Hyun Jo}
\email{hang-hyun.jo@aalto.fi}
\affiliation{BECS, Aalto University School of Science, P.O. Box 12200, Finland}
\author{Raj Kumar Pan}
\email{rajkumar.pan@aalto.fi}
\affiliation{BECS, Aalto University School of Science, P.O. Box 12200, Finland}
\author{Kimmo Kaski}
\email{kimmo.kaski@aalto.fi}
\affiliation{BECS, Aalto University School of Science, P.O. Box 12200, Finland}

\date{\today}

\begin{abstract}
Queuing models provide insight into the temporal inhomogeneity of human dynamics, characterized by the broad distribution of waiting times of individuals performing tasks. We study the queuing model of an agent trying to execute a task of interest, the priority of which may vary with time due to the agent's ``state of mind.'' However, its execution is disrupted by other tasks of random priorities. By considering the priority of the task of interest either decreasing or increasing algebraically in time, we analytically obtain and numerically confirm the bimodal and unimodal waiting time distributions with power-law decaying tails, respectively. These results are also compared to the updating time distribution of papers in the arXiv.org and the processing time distribution of papers in Physical Review journals. Our analysis helps to understand human task execution in a more realistic scenario.
\end{abstract}

\pacs{89.20.-a, 89.75.Da, 89.65.-s}

\maketitle

\section{Introduction}

Understanding human behavior is a fascinating and challenging subject that has been explored extensively in the fields of sociology, economics, and psychology for centuries~\cite{Zipf1949,Becker1978,Ajzen1980}. Recently, with the need to gain deeper insight into complex social systems, the human behavior has also been studied from the perspective of other scientific disciplines including physical and computational sciences~\cite{Castellano2009}. This is because the concepts and methods developed for investigating physical processes have turned out to be applicable to understand the human and social phenomena. Moreover, the huge amounts of digital data have enabled us to explore the human behavior at the very high resolution. The empirical studies based on such dataset have shown that the timing of human actions is not simply random but correlated or bursty~\cite{Barabasi2005,Wu2010,Karsai2011}. Such correlations are often characterized by the heavy tailed or power-law distribution of waiting or response time $\tau$ as $P(\tau)\sim\tau^{-\alpha}$, where $\alpha\approx 1$ for e-mail communication and $1.5$ for letter correspondence, respectively~\cite{Barabasi2005,Oliveira2005}. Here $\tau$ is defined as the time taken by the user to respond to a received message.

The origin of bursts in human dynamics has been explored for years. The long inactive periods at nighttime as well as on weekends are known to result in the bursts of e-mail communication~\cite{Malmgren2008,Malmgren2009a}, where the circadian and weekly cyclic patterns of humans were modeled by the non-homogeneous Poisson process. In addition to these human cyclic patterns, the human dynamics can be also affected by other human factors, such as the task execution behavior. To figure out the role of task execution behavior we have adopted the framework of queuing theory~\cite{Barabasi2005,Vazquez2005,Vazquez2006,Anteneodo2009}, where the to-do list of an agent is modeled as a finite length queue. The agent assigns priorities to tasks in the queue and then executes them according to some protocol. As in the Barab\'asi's model~\cite{Barabasi2005}, if the task with highest priority is executed first, then the one with low priority must wait for a relatively long time in the queue, leading to the heavy tailed distribution of waiting times. Barab\'asi's priority queuing model is based on a few key assumptions: (i) It is a single agent model, where the agent has no explicit interaction with other agents, (ii) all the tasks in the queue are of the same type, such as replying to the received message, and (iii) the priority assigned to each task is kept fixed over time. Several variants of this model have been studied by means of relaxing at least one of these assumptions~\cite{Blanchard2007,Oliveira2009,Min2009,Jo2011}. Here we perform the similar study by relaxing the assumptions (ii) and (iii) and  consider that only one type of the task has the time-varying priority. In the similar context, Blanchard and Hongler studied the queuing systems on the basis of population dynamics, where the individuals in a city (tasks in a queue) are assigned with the priorities increasing with time due to aging or the deadline effect, and finally dying (executed) according to ``the highest priority first'' protocol~\cite{Blanchard2007}. However, instead of the population dynamics based models, the microscopic models with simpler setups might help us to better understand the effect of time-varying priorities on the waiting time distributions.

In this paper, we adopt the microscopic approach and study a single agent queuing model with two tasks: one is the task of interest with time-varying priority and the other is the unexpected and random task with random priority. As an example let us consider a situation in which a researcher wants to complete the research project of current interest, but he or she gets continuously distracted by meetings, coffee breaks and so on. The priority of the task of interest recognized by the researcher can vary due to his or her state of mind, such as mood and happiness. The priority may increase in some cases, such as when the deadline assigned to the task approaches~\cite{Alfi2007}, or decrease in other cases, for example, when the researcher gradually loses interest in that task. For convenience, we call the task of interest and the random task as type-A and type-B tasks, respectively. This model with one type-A and one type-B tasks is minimal and can be generalized to more complex scenarios, such as the case of an agent with many tasks of various types or the case with interacting agents.

The paper is organized as follows. In Section~\ref{sec:model}, we study the models with different cases of time-varying priorities. For each case we obtain the analytic solutions for the waiting time distribution that are confirmed by numerical simulations. In Section~\ref{sec:empirical}, we show how the extended versions of our models to many agents can be used to understand the paper updating mechanism in arXiv.org and the paper reviewing process in Physical Review journals. Finally we summarize the results and make conclusions in Section~\ref{sec:conclusion}.

\begin{figure}[!t]
  \includegraphics[width=\columnwidth]{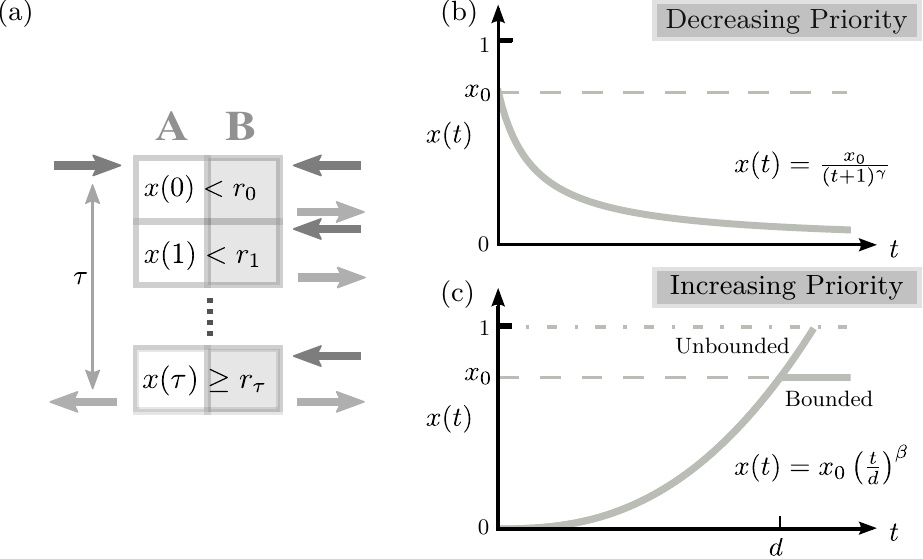}
  \caption{Schematic diagram of time-varying priority queuing model. (a) The task of interest (type-A) has the time-varying priority $x(t)$, which is compared with the random task (type-B) with random number $r_t$ at each time step and executed if $x(t)\geq r_t$. (b) The case with power-law decreasing priority. (c) The case with power-law increasing priority: bounded and unbounded cases. See the text for details.}
  \label{fig:model}
\end{figure}

\section{Model}
\label{sec:model}

We consider an agent with a queue of size $2$ as depicted in Fig.~\ref{fig:model}(a). The first site of the queue is occupied by the task of interest with time-varying priority (type-A task). When the type-A task is introduced to the queue at the time step $t=0$, its priority is given by the random number $x_0$ drawn from the uniform distribution of $(0,1)$ and then it varies with time. The second site of the queue is for the random task (type-B task), which is replaced by a new random task at every time step. The priority of the type-B task introduced at time step $t$ is given by the random number $r_t$ drawn from the uniform distribution of $(0,1)$. At each time step $t$, the priority $x(t)$ is compared with the random number $r_t$ and executed when $x(t)\geq r_t$. Here the execution time defines the waiting time $\tau$ for which the type-A task waits for the execution since it is introduced to the queue. This is the deterministic case where only the highest priority task is executed.

We also consider a stochastic version of the model, where the lower priority task is allowed to be executed with small probability. At each time step, the highest priority task is executed with the probability $p$, while with probability $1-p$ one of two tasks is selected at random for the execution, which we call random selection. This implies that the priority is not the only determinant for the execution. This can also be interpreted as the ``trembling hand'' effect in the game theory~\cite{Weibull1997}. In general, the waiting time distribution $P(\tau)$ is obtained by 
\begin{eqnarray}\label{eq:P_tau}
    P(\tau)&=&\int_0^1 dx_0 \left[
    \prod_{t=0}^{\tau-1}\left\{p[1-x(t)]+\tfrac{1-p}{2}\right\} \right] \left[p x(\tau)+\tfrac{1-p}{2}\right]  \nonumber\\
    &=&p^{\tau+1}\int_0^1 dx_0
    \left[ \prod_{t=0}^{\tau-1}\left[1-x(t)+\epsilon\right] \right] \left[x(\tau)+\epsilon\right],
\end{eqnarray}
where $\epsilon\equiv \frac{1-p}{2p}$ is assumed to be very small in this paper.

\subsection{Fixed priority}

\begin{figure}[!t]
\includegraphics[width=.9\columnwidth]{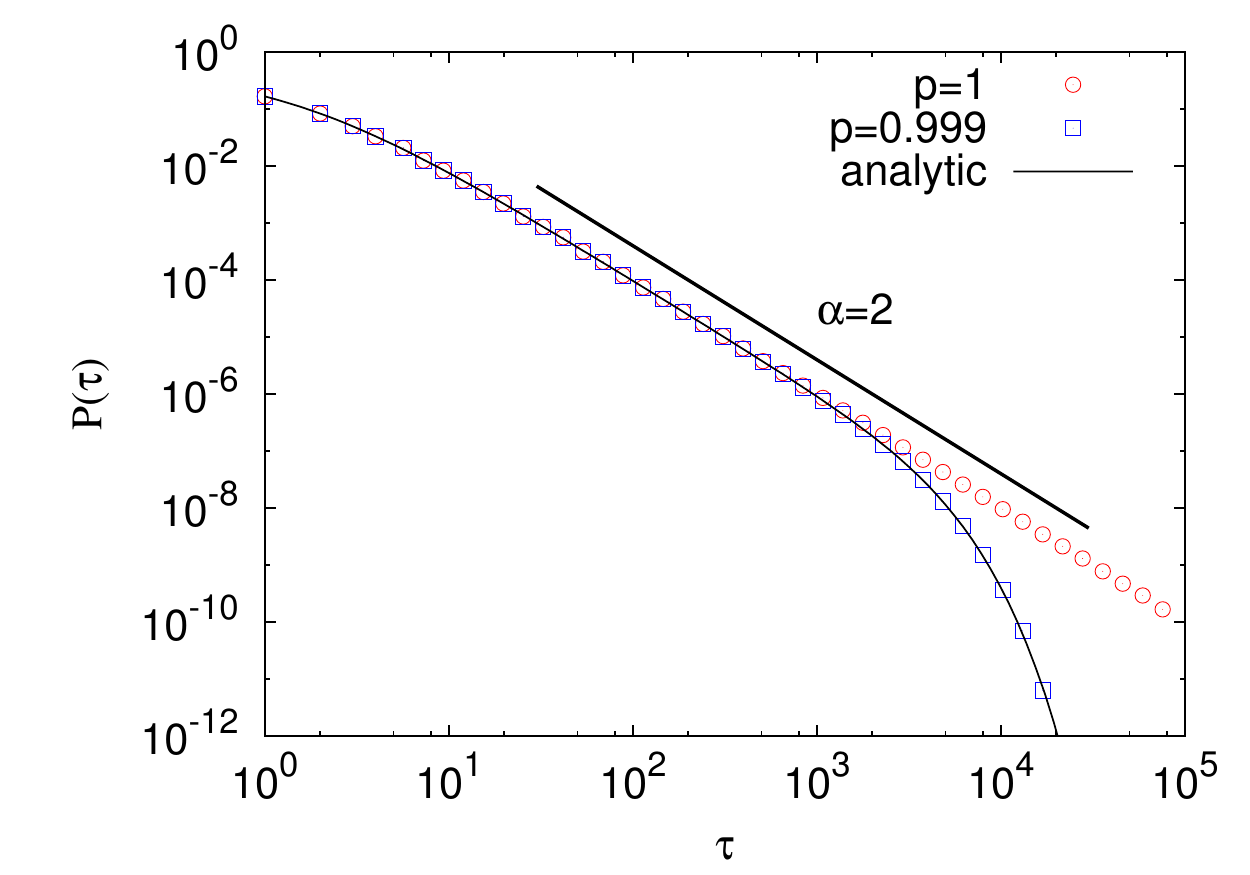}
\caption{The case with fixed priority as $x(t)=x_0$ for deterministic ($p=1$) and stochastic ($p=0.999$) versions. For the numerical results we used $10^9$ samples. The solid line for the stochastic version represents the analytic solution of Eq.~(\ref{eq:fixed_sol}).}
\label{fig:fixed}
\end{figure}

We first consider the case of fixed priority. Here the priority of the type-A task is fixed, i.e. $x(t)=x_0~\forall t$. The waiting time distribution is given by
\begin{eqnarray}\label{eq:fixed_sol}
    P(\tau)&=&p^{\tau+1}\left[\tfrac{(1+\epsilon)^{\tau+2}-\epsilon^{\tau+2}}{(\tau+1)(\tau+2)}
    + \tfrac{\epsilon(1+\epsilon)[(1+\epsilon)^\tau-\epsilon^\tau]}{\tau+1}\right].
\end{eqnarray}
In the limit of $\tau\gg 1$, it can be approximated by
\begin{eqnarray}\label{eq:fixed_sol1}
    P(\tau) &\approx & (\tau^{-2}+2\epsilon\tau^{-1})e^{-\tau/\tau_c(p)},
\end{eqnarray}
where we define the cutoff of distribution as $\tau_c(p)\equiv [\ln(1/p)]^{-1}\approx (2\epsilon)^{-1}$. In the scaling regime, i.e. $1\ll \tau\ll \tau_c\approx (2\epsilon)^{-1}$, the first term $\tau^{-2}$ dominates the distribution function, so that $P(\tau) \approx \tau^{-\alpha}$ with $\alpha=2$. In the deterministic case with $p=1$, $P(\tau)=\frac{1}{(\tau+1)(\tau+2)}\approx\tau^{-2}$ in the limit of $\tau\gg 1$. Note that the small probability of random selection does not change the scaling exponent but leads to the finite cutoff of the distribution. Figure~\ref{fig:fixed} shows that the analytic solution is confirmed by the numerical simulations for both deterministic ($p=1$) and stochastic ($p=0.999$) cases. 

\subsection{Power-law decreasing priority}
\label{subsec:decrease}

Now we consider the case when the priority of the type-A task changes over time. This change might be caused by the internal factors related to the agent as well as to the external factors. We consider the simple case with the algebraically decreasing priority as
\begin{equation}
x(t)=\frac{x_0}{(t+1)^{\gamma}}.
\end{equation}
The decreasing speed of the priority is controlled by the exponent $\gamma$. If $\gamma=0$, the model reduces to the fixed priority case, where the waiting time distribution decays as the power-law with exponent $\alpha=2$. For the larger value of $\gamma$, the priority decays faster so that the type-A task becomes less likely to be executed. Hence one may expect that this will lead to a heavier tail of waiting time distribution and consequently to a smaller value of $\alpha$. 

At first we analyze the case with $\gamma\ll 1$. Since the priority $x(t)$ decreases very slowly, we can use the approximation of $x(t)\approx x(\tau)=\frac{x_0}{(\tau+1)^{\gamma}}$. Putting this into Eq.~(\ref{eq:P_tau}), we get
\begin{eqnarray}
    P(\tau)&\approx& p^{\tau+1}\bigg\{\tfrac{(\tau+1)^{\gamma-1}}{\tau+2} \left[ (1+\epsilon)^{\tau+2}-\left(1+\epsilon-\tfrac{1}{(\tau+1)^\gamma}\right)^{\tau+2}\right]\nonumber\\
    && - (\tau+1)^{\gamma-1}\left[\epsilon+\tfrac{1}{(\tau+1)^\gamma}\right] \left[1+\epsilon-\tfrac{1}{(\tau+1)^\gamma}\right]^{\tau+1}\nonumber\\
    && + (\tau+1)^{\gamma-1}\epsilon(1+\epsilon)^{\tau+1} \bigg\}.\label{eq:smallGamma}
\end{eqnarray}
For large values of $\tau$ and with the fact that $\tau^\gamma$ is close to $1$, the above equation reduces to
\begin{eqnarray}\label{eq:smallGammaApprox}
    P(\tau)&\approx&(\tau^{-(2-\gamma)}+2\epsilon\tau^{-(1-\gamma)})e^{-\tau/\tau_c(p)}.
\end{eqnarray}
In the scaling regime of $1\ll \tau\ll \tau_c\approx (2\epsilon)^{-1}$, the first term $\tau^{-(2-\gamma)}$ dominates the distribution function, so that $P(\tau) \approx \tau^{-\alpha}$ with $\alpha=2-\gamma$. Thus the exponent $\alpha$ is smaller for larger value of $\gamma$, as expected. 

\begin{figure}[!t]
\includegraphics[width=.9\columnwidth]{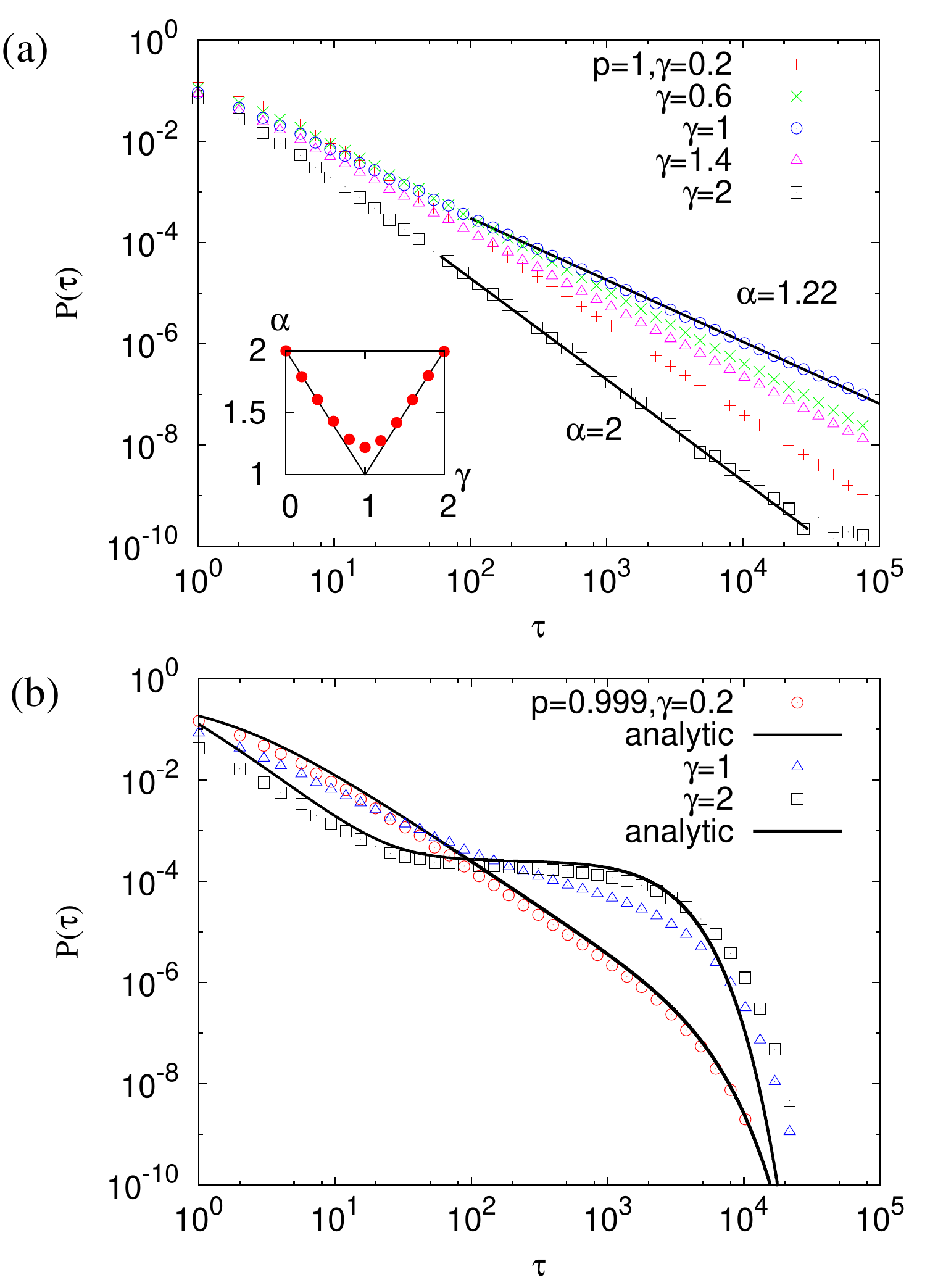}
\caption{The case with power-law decreasing priority as $x(t)=\frac{x_0} {(t+1)^\gamma}$ for deterministic version with $p=1$ (a) and for stochastic version with $p=0.999$ (b). For the numerical results we used up to $10^8$ samples. In the inset of (a) we plot the numerical results of $\alpha(\gamma)$ by the red filled circles with the analytic expectation by the black lines. In (b) the solid lines represent the analytic solutions of Eq.~(\ref{eq:smallGamma}) for $\gamma=0.2$ and Eq.~(\ref{eq:largeGamma}) for $\gamma=2$, respectively.}
\label{fig:decrease}
\end{figure}

For the deterministic version of the model, i.e. $p=1$, we perform the numerical simulations with various values of $\gamma$ as shown in Fig.~\ref{fig:decrease}(a). We find that for any value of $\gamma$ the waiting time distribution follows the power-law behavior with the power-law exponent $\alpha$ as a function of $\gamma$, see the inset of Fig.~\ref{fig:decrease}(a). Our analytic result of $\alpha=2-\gamma$ is numerically confirmed up to $\gamma\approx 0.6$. The numerical results, however, suggest that the value of $\alpha$ has the minimum at $1.22$ for $\gamma=1$, while it increases when $\gamma>1$. In the stochastic case with $p<1$, the second term $\tau^{-(1-\gamma)}$ in Eq.~(\ref{eq:smallGammaApprox}) begins to affect the dominant power-law behavior and the exponential cutoff also becomes finite. This approximate solution is again numerically confirmed for various values of $\gamma$, as depicted in Fig.~\ref{fig:decrease}(b). 

Next, the other limiting case when $\gamma\gg 1$ is considered. Since the value of $x(t)$ decreases very fast, we use the approximation of $x(t)\ll 1$ for all $t$. Thus the product term in Eq.~(\ref{eq:P_tau}) can be expanded in the limit of $\epsilon\ll 1$ as follows
\begin{eqnarray}
    \prod_{t=0}^{\tau-1}[1-x(t)+\epsilon] &\approx& 1-\sum_{t=0}^{\tau-1} [x(t)-\epsilon]\label{eq:expand}\\
    &\approx& 1+\epsilon\tau-x_0\tfrac{1-\tau^{1-\gamma}}{\gamma-1}.
\end{eqnarray}
Then the waiting time distribution is obtained by
\begin{eqnarray}
    P(\tau)&\approx&p^{\tau+1}\bigg\{\tfrac{1}{2(\tau+1)^\gamma} -\tfrac{1-\tau^{1-\gamma}}{3(\gamma-1)(\tau+1)^\gamma}\nonumber\\
    && + \epsilon\left[1-\tfrac{1-\tau^{1-\gamma}}{2(\gamma-1)}+\tfrac{\tau}{2(\tau+1)^\gamma}\right]+\epsilon^2\tau\bigg\}\label{eq:largeGamma}.
\end{eqnarray}
Using the assumption of $\gamma\gg 1$ and the condition that $\epsilon\tau\ll 1$, the above equation can be approximated as
\begin{eqnarray}
    P(\tau)&\approx& \left[\left(\tfrac{1}{2}-\tfrac{1}{3\gamma}\right)\tau^{-\gamma}+\epsilon \left(1-\tfrac{1}{2\gamma}\right)\right]e^{-\tau/\tau_c(p)}.\label{eq:largeGammaFinal}
\end{eqnarray}
For the scaling regime, the waiting time distribution is dominated by the first term, leading to $P(\tau)\approx\tau^{-\alpha}$ with $\alpha=\gamma$. This is numerically confirmed up to $\gamma>1.4$ in the deterministic case as shown in the inset of Fig.~\ref{fig:decrease}(a). For the stochastic case with $p=0.999$, we find that the numerical simulations confirm the analytic solutions but with some deviation due to the approximation we have used, see Fig.~\ref{fig:decrease}(b).

The result of Eq.~(\ref{eq:largeGammaFinal}) can be understood more intuitively as following: For large $\tau$, the probability that the type-A task is not executed becomes of the order of $1$, i.e. $\prod_{t=0}^{\tau-1}[1-x(t)+\epsilon]\approx \mathcal{O}(1)$. Therefore, the only factor relevant to determine $P(\tau)$ is the probability of the task being executed at time step $\tau$. Thus,
\begin{equation}
    P(\tau)\sim [x(\tau)+\epsilon]e^{-\tau/\tau_c} \approx [\tau^{-\gamma}+\epsilon]e^{-\tau/\tau_c},
\end{equation}
which implies $\alpha=\gamma$. For the stochastic case we find that the second term in the result is only a function of $\epsilon$ not coupled to $\tau$. That is, the waiting time distribution does not follow a typical power-law with an exponential cutoff as $\tau^{-\alpha}e^{-\tau/\tau_c}$, but shows a bimodal combination of the power-law and exponential distributions. Similar bimodal distribution has been empirically observed for the inter-event time distributions of Short Messages in mobile phone communication~\cite{Wu2010}.

\subsection{Power-law increasing priority}
\label{subsec:increase}

In this Subsection, we consider the case of the priority of the task increasing as a power-law with time, starting from $x(0)=0$. The priority can increase up to some bounded value, which we call the \emph{bounded} case. Here the priority of type-A task is given by 
\begin{equation}
  x(t)= \begin{cases}
    x_0\left(\frac{t}{d}\right)^\beta & \textrm{if $t<d$,}\\
    x_0 & \textrm{if $t \geq d$,}
  \end{cases}
\label{eq:increase}
\end{equation}
where the scale factor $d$ plays the role of a deadline given to the task. $x_0$ can be interpreted as an intrinsic priority of the task and its value is drawn from the uniform distribution of $(0,1)$. 

On the other hand, the priority of the type-A task can increase up to the maximum possible value of $1$, which we call the \emph{unbounded} case. Once the priority reaches $1$, the task is inevitably executed. In this case, the priority $x(t)$ is given by 
\begin{equation}
  x(t)= \begin{cases}
    x_0\left(\frac{t}{d}\right)^\beta & \textrm{if $t<t_c$,}\\
    1 & \textrm{if $t\geq t_c$,}
  \end{cases}
\label{eq:increase1}
\end{equation}
where $t_c\equiv x_0^{-1/\beta}d$ is defined by the condition $x(t_c)=1$. 

\begin{figure}[!t]
\includegraphics[width=.9\columnwidth]{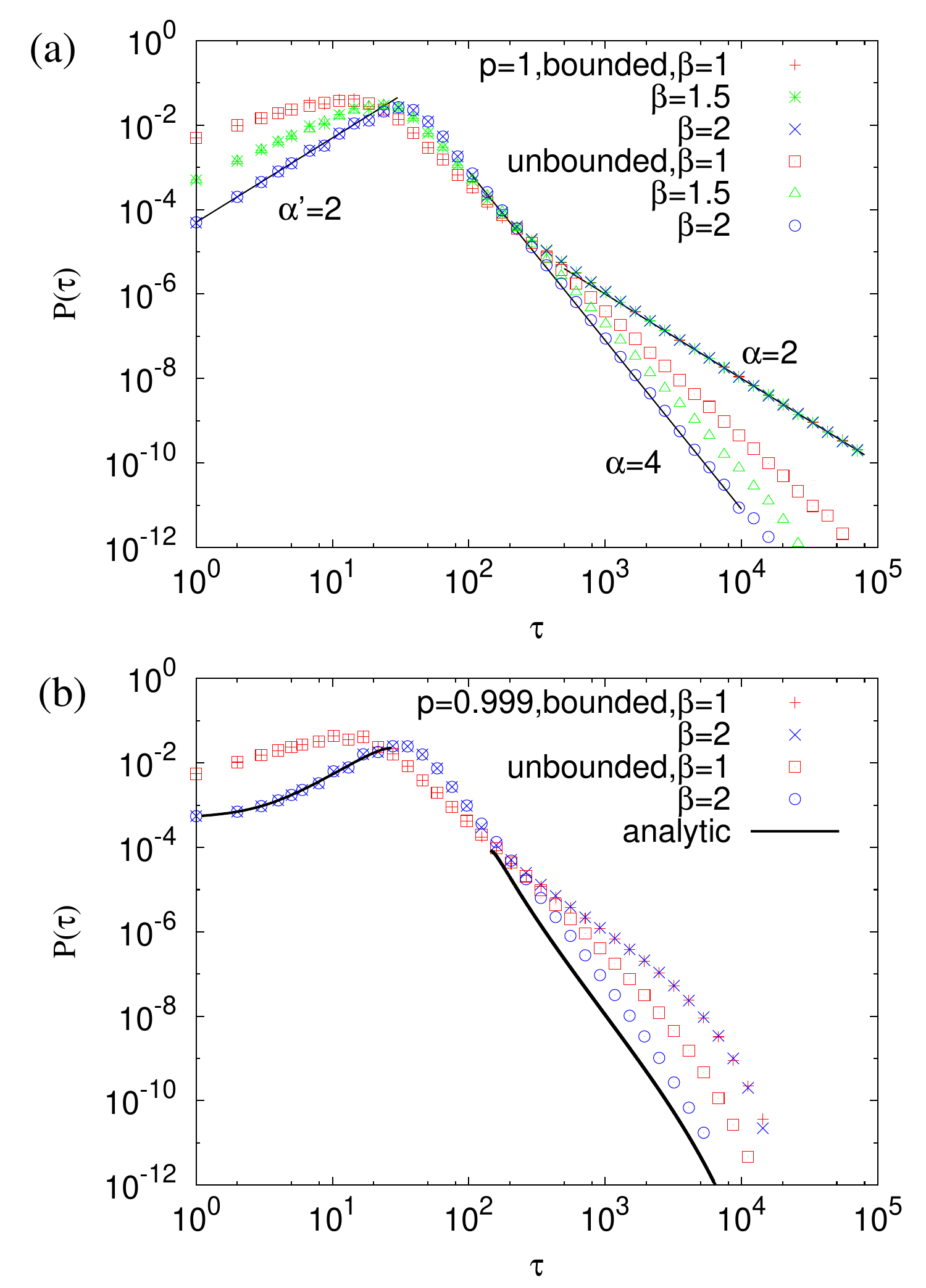}
\caption{The case with power-law increasing priority as $x(t)=x_0 (\tfrac{t}{d})^\beta$ for deterministic version with $p=1$ (a) and for stochastic version with $p=0.999$ (b). The value of $x(t)$ can increase to $x_0$ in the bounded case, where $x(t)=x_0$ for $t\geq d$. On the other hand the value of $x(t)$ can increase to $1$ in the unbounded case. For the numerical results we used up to $10^9$ samples. In (b) the solid lines for $\beta=2$ represent the analytic solutions of Eq.~(\ref{eq:smallTau}) for $\tau<30$ and Eq.~(\ref{eq:largeTau}) for $\tau>150$, respectively.}
\label{fig:increase}
\end{figure}

For both the bounded and unbounded cases, we can adopt the approximation used in Eq.~(\ref{eq:expand}) in the limit of $\tau\ll d$, resulting in
\begin{eqnarray}
    P(\tau) &\approx& p^{\tau+1}\int_0^1 dx_0 \left[1+\epsilon\tau-x_0 \sum_{t=0}^{\tau-1} \left(\tfrac{t}{d}\right)^\beta \right] \left[x_0 \left(\tfrac{\tau}{d}\right)^\beta+\epsilon\right] \nonumber\\
    &\approx& p^{\tau+1}\big[ \tfrac{1}{2}(\tfrac{\tau}{d})^\beta + 
    \tfrac{\epsilon d}{2}(\tfrac{\tau}{d})^{\beta+1} 
    -\tfrac{\epsilon d}{2(\beta+1)}(\tfrac{\tau-1}{d})^{\beta+1} \nonumber\\
    && -\tfrac{d}{3(\beta+1)} (\tfrac{\tau}{d})^{\beta} (\tfrac{\tau-1}{d})^{\beta+1}
    +\epsilon(1+\epsilon\tau)\big].\label{eq:smallTau}
\end{eqnarray}
In the limit of $\epsilon\tau\ll 1$, the above equation can be approximated as 
\begin{eqnarray}
    P(\tau) &\approx&\left[\tfrac{1}{2}(\tfrac{\tau}{d})^{\beta}+\epsilon\right]e^{-\tau/\tau_c(p)}.
\end{eqnarray}
For the range of $\epsilon^{1/\beta}\ll \tfrac{\tau}{d}\ll 1$, the waiting time distribution is dominated by the first term, indicating a power-law increasing behavior as $P(\tau)\sim \tau^{\alpha'}$ with $\alpha'=\beta$. This solution is compared to the numerical results for both the deterministic and stochastic versions of the model, where we set $d=100$, see Fig.~\ref{fig:increase}.

On the other hand, in the limit of $\tau \gg d$, the waiting time distributions decay as a power-law, but with different tail exponents for bounded and unbounded cases. In the bounded case, when $t\geq d$, the priority is fixed as $x(t)=x_0$, corresponding to the fixed priority model. Therefore, we get $\alpha=2$, which is independent of the values of $\beta$ and $d$. To obtain the distribution for the unbounded case, we use the approximations that $x(t)\ll 1$ if $t<d$ and that $x(t)\approx x(\tau)$ if $t\geq d$. In addition, the type-A task always gets executed as the priority reaches $1$, implying that $x(\tau)+\epsilon\leq 1$. This means that the value of $x_0$ has an upper bound of $x_c$ given by $x_c (\tfrac{\tau}{d})^\beta+\epsilon=1$. Thus the upper limit of the integral range in Eq.~(\ref{eq:P_tau}) is also limited to $x_c$, leading to
\begin{eqnarray}
    P(\tau) &\approx& p^{\tau+1}\int_0^{x_c} dx_0 \left[1+d\epsilon-x_0
    \sum_{t=0}^{d-1} \left(\tfrac{t}{d}\right)^\beta \right] \nonumber \\ 
    && \times \left[ 1-x_0(\tfrac{\tau}{d})^\beta+\epsilon \right]^{\tau-d} \left[ x_0 \left(\tfrac{\tau}{d}\right)^\beta+\epsilon \right] \label{eq:largeTau}\nonumber\\
    &=&
    \tfrac{(1+\epsilon)^{\tau-d+1}\epsilon(1+d\epsilon)-(2\epsilon)^{\tau-d+1}\left[1+d\epsilon
    -A(1-\epsilon)/a_\tau\right]}{a_\tau(\tau-d+1)} \nonumber \\
    && + \tfrac{(1+\epsilon)^{\tau-d+2}\left[(1+d\epsilon)a_\tau-A\epsilon\right]-(2\epsilon)^{\tau-d+2}\left[(1+d\epsilon)a_\tau-A(2-\epsilon)\right]}{a^2_\tau (\tau-d+1)(\tau-d+2)}\nonumber \\
    && - \tfrac{2A\left[(1+\epsilon)^{\tau-d+3}-(2\epsilon)^{\tau-d+3}\right]}{a^2_\tau(\tau-d+1)(\tau-d+2)(\tau-d+3)},
\end{eqnarray}
where $a_\tau\equiv (\tfrac{\tau}{d})^\beta$ and $A\equiv\tfrac{d}{\beta+1}(\tfrac{d-1}{d})^{\beta+1}$. The above equation can be approximated in the limit of $\tau\gg 1$ as 
\begin{eqnarray}
    P(\tau) &\approx& d^\beta(\tau^{-(\beta+2)}+2\epsilon \tau^{-(\beta+1)})e^{-\tau/\tau_c(p)}.
\end{eqnarray}
In the scaling regime, i.e. $\epsilon\tau\ll 1$, we find that $P(\tau) \sim\tau^{-\alpha}$ with $\alpha=\beta+2$. We performed the numerical simulations with various values of $\beta$ for both deterministic and stochastic cases. In Fig.~\ref{fig:increase} we find that the simulations confirm the analytic solutions for both deterministic and stochastic cases for the range of $\tau<d$. The range of $\tau>d$ in the stochastic case is only qualitatively matched, which is again due to the approximations we have used.

\section{Empirical Analysis}
\label{sec:empirical}

The results of the models for different cases of time-varying priorities in Section~\ref{sec:model} can be applied to understand the empirical distributions of the updating times of papers that have appeared in arXiv.org (arXiv in short) and of the processing times of papers published in Physical Review journals (PR in short). The essential ingredients of the empirical setup include the following: (i) An agent has a given task of interest, the execution of which gets postponed due to the other tasks and (ii) the priority of the task varies with time.

\subsection{Description of the datasets}

The first dataset consists of all the papers that appeared on arXiv between January 1995 and March 2010. For each paper we consider its \emph{updating time}, defined as the time interval in days between submission and the final update. The papers that were not updated have been discarded in our analysis. This dataset contains 185151 papers. We divide this dataset into three groups depending on the number of authors in each paper, denoted by $n$. The groups of $n=1$, $n=2$, and $n\geq 3$ consist of 55382, 56572, and 73197 papers, respectively. The papers with more authors have the smaller average updating time: 220.8, 188.5, and 160.0 days for groups of $n=1$, $n=2$, and $n\geq 3$.

The second dataset consists of papers that appeared in PR between January 2000 and December 2009. For each paper we consider its \emph{processing time}, defined as the time interval in days between submission and publication. We have retained only the research articles and have removed articles of other types such as errata and publisher's notes, which often have different processing times. The exact publication dates of papers printed before 2000 were not available and hence were ignored. This dataset consists of 157484 papers. We divide it into two 5-year periods, i.e. sets of papers appearing for 2000-2004 and for 2005-2009, respectively. The average updating time decreased from $200.3$ days for 2000-2004 to $163.9$ days for 2005-2009. We would like to note that many researchers submit their papers to arXiv at the same time while submitting them to journals and update the arXiv versions on acceptance of the article by the journal. Both the updating time and the processing time are denoted by $\tau$.

\begin{figure}[!t]
\includegraphics[width=.9\columnwidth]{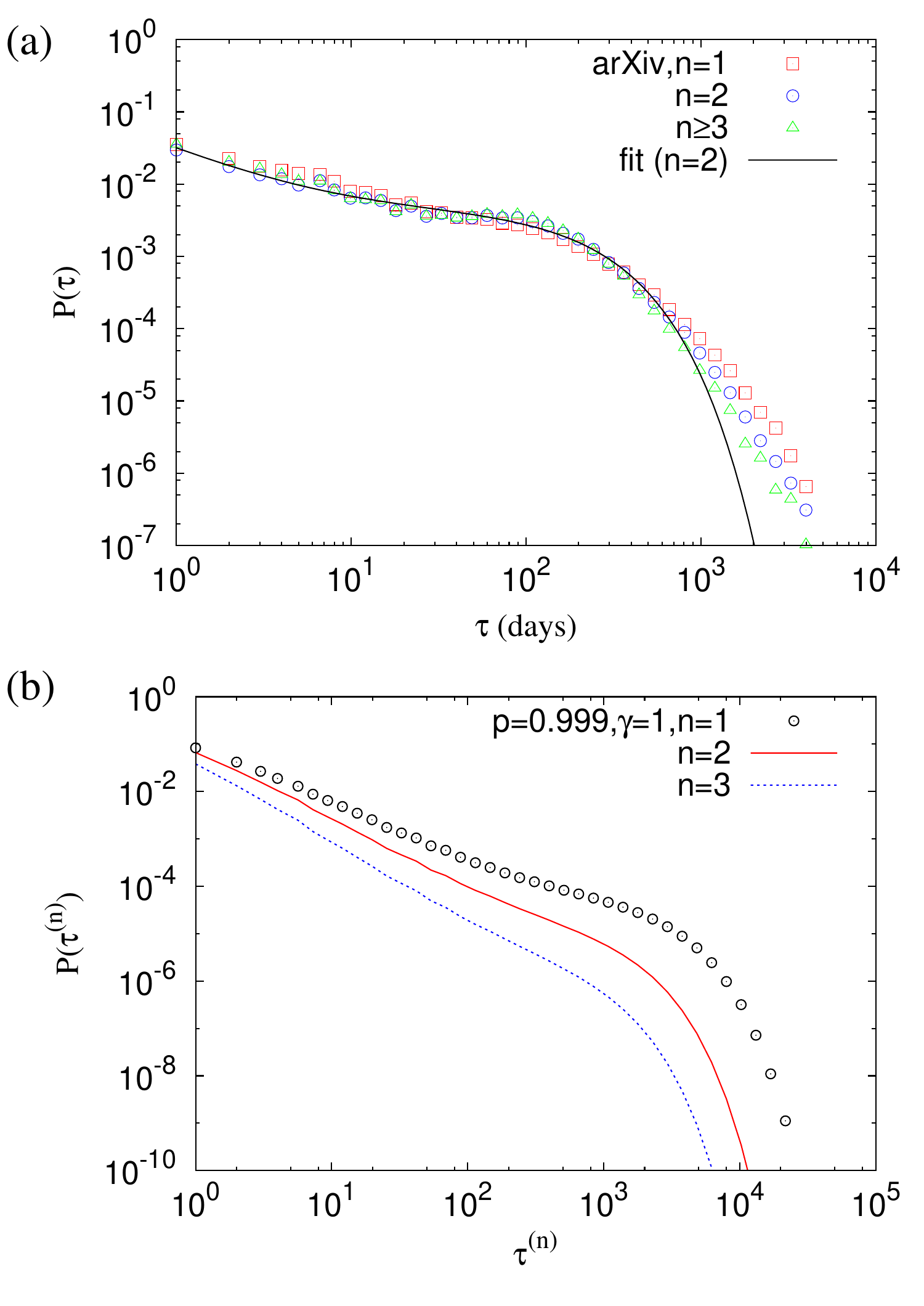}
\caption{(a) Probability distributions of updating times for papers by single authors ($n=1$), two authors ($n=2$), and more than two authors ($n\geq 3$) in arXiv.org. The solid line indicates the bimodal curve $(\tau^{-\alpha}+c)e^{-\tau/\tau_c}$ fitting to the group of $n=2$. The estimated values of parameters are provided in the text. (b) Waiting time distributions of the stochastic model with power-law decreasing priority in Subsection~\ref{subsec:decrease} and its extended versions to $n$ agents. Here $\tau^{(n)}$ is defined as $\min\{\tau_i\}_{i=1,\cdots,n}$ with each $\tau_i$ drawn from the distribution of the single agent model.}
\label{fig:empiricalArxiv}
\end{figure}

\begin{figure}[!t]
\includegraphics[width=.9\columnwidth]{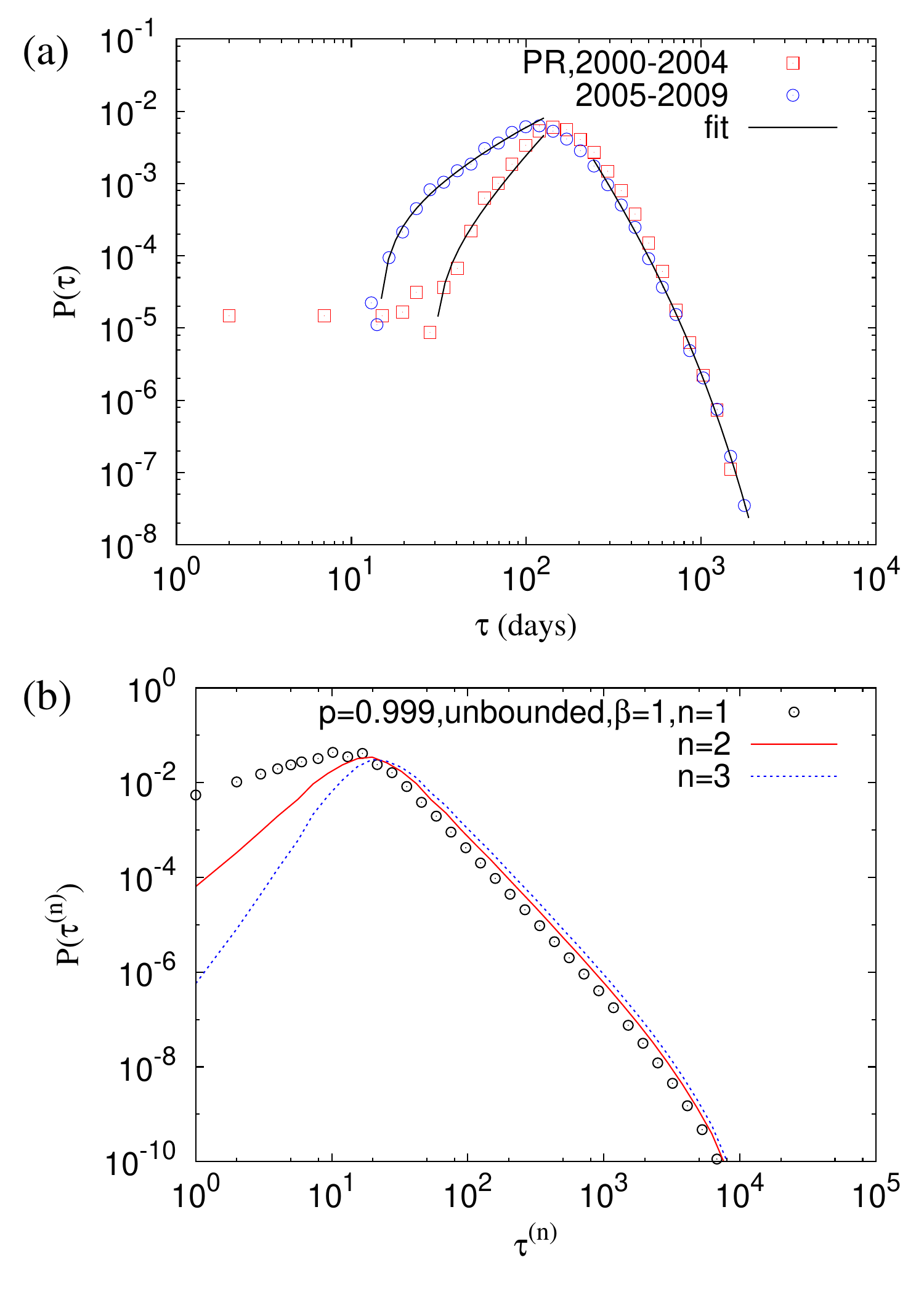}
\caption{(a) Probability distributions of processing times in Physical Review journals for different periods. We fit the increasing part and the tail part of the distributions with the curves $(\tau^{\alpha'}+c')e^{-\tau/\tau_c}$ and $\tau^{-\alpha}e^{-\tau/\tau_c}$, respectively. The fitting curve for the tail of 2000-2004 is not shown for clear presentation. The estimated values of parameters are provided in the text. (b) Waiting time distributions of the stochastic model with power-law increasing priority in Subsection~\ref{subsec:increase} and its extended versions to $n$ agents. Here $\tau^{(n)}$ is defined as $\max\{\tau_i\}_{i=1,\cdots,n}$ with each $\tau_i$ drawn from the distribution of the single agent model.}
\label{fig:empiricalPR}
\end{figure}

\subsection{Updating time distributions in arXiv.org}

We plot the updating time distributions of arXiv papers for groups of $n=1$, $n=2$, and $n\geq 3$, as depicted in Fig.~\ref{fig:empiricalArxiv}(a). In all cases, the initial part of the distribution ($\tau<50$) shows a power-law decaying behavior, which is followed by an exponentially decaying tail. The distribution also shows a hump around $\tau=100$ indicating its bimodal nature. These distributions can be fitted by a simplified bimodal curve of the form $P(\tau) \propto (\tau^{-\alpha}+c)e^{-\tau/\tau_c}$. We ignored very large updating times ($\tau>2000$) due to the inconclusive fitting results. The parameter values are estimated as $\alpha \approx 0.76(7)$ and $\tau_c \approx 234(29)$ for $n=1$ group, $\alpha \approx 0.99(13)$ and $\tau_c \approx 189(17)$ for $n=2$ group, and $\alpha \approx 1.16(15)$ and $\tau_c \approx 166(7)$ for $n\geq 3$ group, respectively. The fitted curve for $n=2$ group is plotted in Fig.~\ref{fig:empiricalArxiv}(a). It is found that the value of $\alpha$ is slightly larger for the papers with more authors. The cutoff updating times $\tau_c$ are consistent with the corresponding average updating times within error bars.

Let us consider the updating mechanism of the papers submitted to arXiv. Immediately after submission the authors are most alert and are likely to find some anomalies in their paper and subsequently update it. As time passes, they get more involved with other projects and the probability to find irregularities decreases, leading to a decreasing priority of the updating task. In addition, there is a finite probability that the authors might receive a referee report, and hence will update the paper with a revised version even when the priority of the updating task is smaller than the other tasks. Thus, the decreasing priority and the stochastic selection of the updating task can qualitatively describe the bimodal updating time distributions in arXiv. However, for the papers by more than one author the interaction among authors is also an important factor for determining the updating time. In this case, we consider the generalized model with $n$ agents. One of the authors who first finds the anomalies in the paper is assumed to update the paper. Then, the resultant updating time is determined by the shortest updating time as $\tau^{(n)}\equiv\min\{\tau_i\}_{i=1,\cdots,n}$, where each $\tau_i$ is drawn from the distribution of the single agent model in Subsection~\ref{subsec:decrease}. The distribution of $\tau^{(n)}$ reads
\begin{equation}
    P(\tau^{(n)})=c_nP(\tau_1=\tau^{(n)})\prod_{i=2}^n P(\tau_i\geq \tau^{(n)}),
\end{equation}
where $c_n$ is the normalization constant. We numerically obtain $P(\tau^{(n)})$ from the numerical results of $P(\tau_i)$ of the single agent model for small values of $n$, as shown in Fig.~\ref{fig:empiricalArxiv}(b). The decreasing average updating time according to $n$ is observed as in the case of empirical results. The power-law exponent for the range of small updating times, denoted by $\alpha_{\rm min}$, tends to increase with $n$. This tendency is also consistent with the empirical observation apart from the exact values, while the discrepancy in the values of power-law exponent would require some additional or different approach beyond the current version of our models.

\subsection{Processing time distributions in Physical Review journals}

Next, we plot the processing time distributions of PR papers for periods 2000-2004 and 2005-2009, as depicted in Fig.~\ref{fig:empiricalPR}(a). In both cases, when compared to the bimodality found in the arXiv case, the distributions are unimodal with peaks at $\tau\approx 150$ and $120$ for 2000-2004 and 2005-2009, respectively. We fit the tail of distribution, by using the functional form of $P(\tau)\propto \tau^{-\alpha}e^{-\tau/\tau_c}$. We estimate $\alpha\approx 2.7(3)$ and $\tau_c \approx 234(27)$ for 2000-2004 and $\alpha\approx 3.2(3)$ and $\tau_c \approx 337(38)$ for 2005-2009, respectively. While the values of $\alpha$ are the same within error bars, the value of $\tau_c$ increases from 2000-2004 to 2005-2009. Then we fit the increasing part of distribution with the form $P(\tau)\propto (\tau^{\alpha'}+c')e^{-\tau/\tau_c}$, where we have assumed the same value of $\tau_c$ as for the tail. The estimated values are $\alpha'\approx 3.2(2)$ for 2000-2004 and $\alpha'\approx 1.6(1)$ for 2005-2009, respectively. The fitted curves are plotted in Fig.~\ref{fig:empiricalPR}(a). 

The publication process of PR can be understood as a simple queuing process from the point of view of the referee. Among many factors affecting the processing of the paper, we consider only the time taken by the referee as the main factor. The other factors, such as the time taken by editors, are considered to be fixed and hence they are expected not to change the shape of the distribution but only to affect the location of the peak. The referee assigns some initial priority to the received paper, which may depend upon the quality of the paper and the referee as well. The priority of the reviewing task increases with time, as the journal editor hurries up the referee in case when he or she does not submit the review report within a stipulated period. The referees may also select the reviewing task randomly, independent of the priority. Hence the simple mechanism of the power-law increasing priority of the referee can describe the unimodal processing time distributions in the case of PR. As the number of referees per paper varies, we can consider the generalized model with $n$ independent referees. The processing time of a paper is determined by the longest reviewing time, i.e. $\tau^{(n)}\equiv \max\{\tau_i\}_{i=1,\cdots,n}$. Here each $\tau_i$ is drawn from the distribution of the single agent model in Subsection~\ref{subsec:increase}. The distribution of $P(\tau^{(n)})$ reads
\begin{equation}
    P(\tau^{(n)})=c_nP(\tau_1=\tau^{(n)})\prod_{i=2}^n P(\tau_i\leq \tau^{(n)}),
\end{equation}
where $c_n$ is the normalization constant. When $\tau^{(n)}\ll d$, since $P(\tau_i)\sim \tau_i^{\alpha'}$ for each $i$, the leading term of $P(\tau^{(n)})$ becomes $\tau^{(n)\alpha'_{\rm max}}$ with $\alpha'_{\rm max}=n\alpha'+n-1$. On the other hand, if $\tau^{(n)}\gg d$, $P(\tau_i\leq \tau^{(n)})$ is of the order of $1$ due to the unimodality of the distribution. Thus, the tail of the distribution is described by the same exponent as in the single agent model, i.e.  $\alpha_{\rm max}=\alpha$, which is independent of $n$. Then, by means of $\alpha=\alpha'+2$ obtained in the unbounded case of the single agent model, we get the following equation:
\begin{equation}
    n=\frac{\alpha'_{\rm max}+1}{\alpha_{\rm max}-1}.
\end{equation}
Since the information about the numbers of referees in PR is not accessible, we infer the \textit{effective} number of referees by plugging the empirical values of power-law exponents $\alpha$ and $\alpha'$ into $\alpha_{\rm max}$ and $\alpha'_{\rm max}$, respectively. We obtain $n\approx 2.5$ for 2000-2004 and $n\approx 1.2$ for 2005-2009, respectively. The decreasing value of $n$ is also consistent with the decreasing average processing time, in a sense that the longest reviewing time will be shorter in case of the fewer referees. The additional factors, such as the number of revisions needed and the time taken by the authors for revision, can be considered in an extended version of the model.  

\section{Conclusion}
\label{sec:conclusion}

Various forms of waiting time distributions have been observed in the human dynamics, including the bimodal and the unimodal distributions with power-law or exponentially decaying tails. To figure out the origin of various types of bursty correlations, we have studied a single agent queuing model with two tasks: one is the task of interest whose priority algebraically increases or decreases with time, and the other is the random task with the random priority. If we choose the model to be deterministic, only the highest priority task will be executed. In the stochastic version, however, one of the tasks in the queue is randomly selected for execution with small probability. In all the cases we obtain the approximate analytic solutions for the waiting time distributions of the task of interest, which are confirmed by the extensive numerical simulations. 

For the case of power-law decreasing priority with exponent $\gamma$, we find that the power-law exponent of the waiting time distribution $P(\tau)\sim\tau^{-\alpha}$ shows the non-monotonous behavior as $\gamma$ increases, i.e. $\alpha=2-\gamma$ for $\gamma\ll 1$ and $\alpha=\gamma$ for $\gamma\gg 1$. The stochastic case of the model leads to both the bimodality and the finite cutoff of distributions. For the case of power-law increasing priority with exponent $\beta$, the resulting waiting time distributions increase as power-law $P(\tau)\sim\tau^\beta$ up to some scale factor. The power-law exponent $\alpha$ of tails turns out to be $2$ and $\beta+2$ for the bounded and unbounded case, respectively. The extended versions of our models to many agents are also compared to the empirical distributions of updating times for the papers appeared in arXiv.org (arXiv) and of processing times for the papers published in the Physical Review journals (PR). The updating process in arXiv can be understood as a model where each author's priority of updating the paper decreases with time as a power-law. The processing time of a paper in PR can be understood by means of the reviewing process, where each referee's priority of reviewing the paper increases with time.

Compared to the Barab\'asi's queuing model~\cite{Barabasi2005}, in our models the tasks in the queue are assumed to be of different types. The task of interest among them has the time-varying priority, while its execution is affected by the random tasks. Our simplified models for the human dynamics can be extended or generalized to the more complicated and realistic cases. As an example, a researcher working on two ongoing projects with different time-varying priorities can be studied by means of the model with two tasks of interest and one random task in his or her to-do list.

\begin{acknowledgments}
Financial support by Aalto University postdoctoral program (HJ), from EU's FP7 FET-Open to ICTeCollective Project No. 238597 (KK), and by the Academy of Finland, the Finnish Center of Excellence program 2006-2011, Project No. 129670 (RKP, KK) are gratefully acknowledged.
\end{acknowledgments}


%

\end{document}